# Nanotubes Motion on Layered Materials: A Registry Perspective

Inbal Oz,[1,2] Itai Leven,[1,2] Yaron Itkin,[1] Asaf Buchwalter,[1] Katherine Akulov,[1] Oded Hod[1,2]*

E-mail: odedhod@tau.ac.il

[1]Department of Physical Chemistry, School of Chemistry, The Raymond and Beverly Sackler Faculty of Exact Sciences, Tel Aviv University, Tel Aviv, IL 6997801

[2]The Sackler Center for Computational Molecular and Materials Science, Tel Aviv University, Tel Aviv, IL 6997801

**Abstract**

At dry and clean material junctions of rigid materials the corrugation of the sliding energy landscape is dominated by variations of Pauli repulsions. These occur when electron clouds centered around atoms in adjacent layers overlap as they slide across each other. In such cases there exists a direct relation between interfacial surface (in)commensurability and superlubricity, a frictionless and wearless tribological state. The Registry Index is a purely geometrical parameter that quantifies the degree of interlayer commensurability, thus providing a simple and intuitive method for the prediction of sliding energy landscapes at rigid material interfaces. In the present study, we extend the applicability of the Registry Index to non-parallel surfaces, using a model system of nanotubes motion on flat hexagonal materials. Our method successfully reproduces sliding energy landscapes of carbon nanotubes on Graphene calculated using a Lennard-Jones type and the Kolmogorov-Crespi interlayer potentials. Furthermore, it captures the sliding energy corrugation of a boron nitride nanotube on hexagonal boron nitride calculated using the $h$-BN ILP. Finally, we use the Registry Index to predict the sliding energy landscapes of the heterogeneous junctions of a carbon nanotubes on hexagonal boron nitride and of boron nitride nanotubes on graphene that are shown to exhibit a significantly reduced corrugation. For such rigid interfaces this is expected to be manifested by superlubric motion.

# Introduction

Nano-electromechanical systems (NEMS) present the ultimate miniaturization of electro-mechanical devices[1,2]. Their realization has paved the path for the design of molecular scale devices with unique properties and functionality[3,4]. Nanotubes (NTs) have often been suggested to serve as active components in such systems due to their cylindrical geometry and remarkable mechanical and electronic properties[5–9]. Such setups often involve junctions of NTs and atomically flat surfaces where the detailed lattice structure at the interface determines its tribological properties[10–12]. Gaining a clear understanding of the NT-surface interactions has thus been the focus of several recent computational studies exhibiting the importance of a full atomic-scale description[13]. The tool of choice in such studies is often classical mechanics simulations based on dedicated force-fields that are designed to reproduce the properties of specific junctions as obtained either experimentally or via higher-accuracy computational methods[14–16]. These provided important insights regarding the interplay between lattice commensurability and preferred NT orientations as well as the different mechanism underlying various types of motion including sliding, rolling, and spinning. While such descriptions are highly valuable for the interpretation of experimental observations and the prediction of new phenomena, they may turn computationally demanding with increased force-field sophistication and system dimensions and tend to blur the atomic-scale origin of tribological phenomena in nanoscale junctions.

Recently, an alternative that quantifies the interlayer registry in rigid interfaces has been proposed for modeling the interlayer sliding energy surfaces of a variety of hexagonal layered materials including graphene[17], hexagonal boron nitride ($h$-BN)[18], molybdenum disulfide,[19] and multi-layered nanotubes thereof[7,20,21]. Within this approach, one defines a registry index (RI) as a geometrical parameter that gives a quantitative measure of the degree of commensurability between two lattices. The method focuses on describing the repulsive Pauli interactions that dictate the potential energy landscape for sliding in these systems via simple circle overlap calculations. Hence, it provides a clear and intuitive description of the origin of sliding energy corrugation with negligible computational cost.

Thus far, the RI was successfully applied to parallel surfaces either flat[17–19,22–24] or curved[7,20]. Here, we extend its applicability to treat non-parallel surfaces. As a model system we choose to study the motion of NTs on flat surfaces of hexagonal layered materials. By defining an atom dependent two-dimensional (2D) Gaussian function (replacing the original atomic centered circles) we mimic the effect of reduced Pauli repulsions with increasing distance between atoms belonging to the NT and to the surface. This allows us to fully reproduce the potential energy variations during carbon NT (CNT) spinning, rolling, and sliding on graphene[13] and the sliding motion of a boron nitride nanotube (BNNTs) on $h$-BN as obtained using elaborate force-field calculations. Finally, we use our approach to predict the tribological properties of the heterogeneous interfaces formed between CNTs and $h$-BN or BNNTs and graphene.

# Computational Methods

In the original RI approach each lattice center was assigned a circle of radius $r_i$ that depended on the atomic identity. In the present implementation, in order to obtain smoother and more physical registry index surfaces, we replace these circles by atomic centered 2D Gaussian functions whose standard deviations relate to the original circle radii via $\sigma_i = \gamma r_i$, where $\gamma$ is chosen to reproduce the sliding energy RI landscapes obtained using the original circle-based definition and typically assumes a value of $\gamma = 0.75$ (see supplementary material). Projected Gaussian overlaps between atomic centers belonging to adjacent surfaces are then analytically calculated to evaluate the local degree of repulsive interactions. A simple formula involving sums and differences of the local overlaps is used to define a numerical parameter aimed to quantify the overall inter–facial registry mismatch between the two lattices. In order to generalize the definition of the RI to non-parallel surfaces one should take into account the inter-site distance dependence of the repulsive interactions. To this end, we scale each pair overlap contribution according to the relative distance between the corresponding sites. This follows the spirit of our RI treatment of planar $2H$-$MoS_2$ where the circle radii were chosen to reflect the distance between the relevant pair of sub-layers[19]. To demonstrate this, we start by considering the homogeneous junction of a CNT on graphene. First, we assign to each atomic position within the tube and the surface a 2D Gaussian parallel to the graphene surface. The Gaussian standard deviation is chosen to be $\sigma_C = 0.75 \times r_C = 0.75 \times (0.5 L_{CC})$, where $L_{CC} = 1.42$ Å is the covalent inter-carbon bond in graphene. Next, the projected overlaps between the Gaussians of the tube and those of the surface (see Fig. 1(c)) are calculated according to

$$S_{C_t^i C_g^j} = f(h) \times \int_{-\infty}^{\infty} \int_{-\infty}^{\infty} \mathrm{e}^{-\frac{(\mathbf{r}-\mathbf{r}_i)^2}{2\sigma_i^2}} \mathrm{e}^{-\frac{\left(\mathbf{r}-\mathbf{r}_j\right)^2}{2\sigma_j^2}} dxdy = f(h) \times \frac{\pi \sigma_i^2 \sigma_j^2}{\sqrt{\sigma_i^2 + \sigma_j^2}} \times \mathrm{e}^{-\frac{\left|\mathbf{r}_i - \mathbf{r}_j\right|^2}{2\left(\sigma_i^2 + \sigma_j^2\right)}}, \tag{1}$$

where $\mathbf{r}_i$ and $\mathbf{r}_j$ are 2D vectors representing the projected positions of atoms $C_t^i$ of the tube and $C_g^j$ of the graphene surface on the $XY$ plane parallel to the graphene surface, respectively, and $f(h) = H(R-h) \times \exp[-\alpha_g (h - h_g)]$ is a dimensionless scaling factor serving to effectively reduce the inter-atomic overlap contribution with increasing vertical distance, $h$. With this definition $f(h)$ obtains the value of 1 when the vertical distance equals the equilibrium graphene interlayer distance of $h_g = 3.33$ Å and decays exponentially with a (system dependent) factor of $\alpha_g = 3.0$ Å$^{-1}$ set to reproduce reference data[25]. The Heaviside step function, $H(R-h)$, serves to cut-off all overlap contributions of atoms of the upper tube section, which are assumed to be screened from the surface by the lower tube section. The RI is defined to be proportional to the total overlap area obtained by summing all atomic pair overlaps, $S_{CC}^{tot} = \sum_{i=1}^{N_t} \sum_{j=1}^{N_g} S_{C_t^i C_g^j}$ and normalizes to the range $[0,1]$ similar to its planar system definition[17]:

$$RI_{\text{graphitic}} = \frac{S_{CC}^{tot} - S_{CC}^{AB}}{S_{CC}^{AA} - S_{CC}^{AB}}. \tag{2}$$

To calculate the overlap value at the optimal ($S_{CC}^{AB}$) stacking mode we position the tube on the graphene surface such that its translational vector forms an angle of ($\frac{\pi}{6} - \theta$), $\theta$ being the chiral angle of the tube, with the zigzag graphene direction and the lower hexagon stripe of the tube forms an AB (Bernal) stacking configuration with the graphene surface (see figure 1). For the worst stacking mode overlap ($S_{CC}^{AA}$) we merely shift the tube in the armchair direction by $-L_{CC}$.

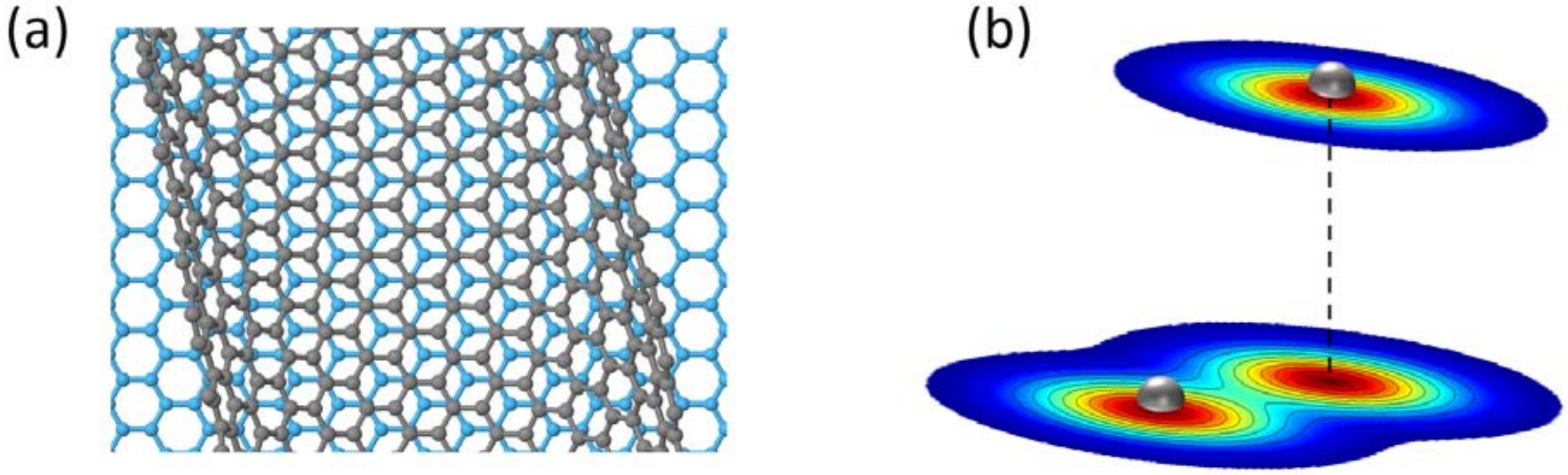

Figure 1: (a) Schematic representation of the optimal configuration of a (20, 10) carbon nanotube on a graphene surface. The lower hexagon stripe of the tube is AB stacked with the underlying flat hexagonal lattice. For clarity, the graphene carbon atoms are colored in cyan and only the lower half of the tube is presented. (b) Illustration of the projected overlap between two 2D Gaussian functions one associated with an atom of the tube and another with an atom of the graphene surface.

A similar procedure is used for BNNTs on $h$-BN and for the heterogeneous junctions of CNTs on $h$-BN and BNNTs on graphene. For these systems, as well, the Gaussian standard deviations are obtained from the circle radii of the original planar RI definitions with $\gamma = 0.75$ and the corresponding optimal and worst planar stacking modes are used for normalization[18,22]. Specifically, for the heterogeneous junctions normalization procedure, stretched unrolled bilayers with matching lattice constants are used[22].

## Results and Discussion

We first demonstrate the performance of the suggested approach by showing that it can successfully reproduce the energy landscapes for various types of CNT motion on flat graphene surfaces calcu-lated using classical force-fields. In Fig. 2a we compare the energy variations recorded during the spinning of $(10, 10)$, $(30, 0)$, and $(20, 10)$ CNTs on a graphene surface as obtained by Buldum et al. using a Lennard-Jones (LJ) type potential[13] and the corresponding RI changes. As can be seen, for all three systems considered the RI calculation fully reproduces the different force-field results down to fine details with a negligible computational cost. Similarly, when considering the sliding and rolling motions of the $(20, 10)$ CNT on graphene (Fig. 2b) excellent agreement is achieved between the calculated force-field energy variations and the corresponding RI changes. Importantly, since we use the same RI parameters and normalization scheme for all motion types a single scaling factor is sufficient to relate the RI and the force-field results. This excellent agreement achieved between the two types of calculations results from the fact that the dominating interactions determining the calculated energy landscape are of short range repulsive nature and hence can be readily captured by the simple RI picture even when the surfaces are not parallel.

To further evaluate the robustness of the RI method for non-parallel surfaces, we repeated the force-field calculations using the anisotropic Kolmogorov-Crespi (KC) potential that was shown to be superior over LJ type expression in describing the intrelayer interactions in graphitic systems[15]. Fig. 3a presents the corresponding energy variations (black line) during the sliding motion of a $(20, 10)$ CNT on a graphene surface. We note that the energy variations are normalized to the number of interacting atoms defined as $N \equiv 2\sum_{i=1}^{Ntube} f(h_i)$ where $h_i$ is the vertical height of atom $i$ of the tube above the graphene surface, $N_{tube}$ is the total number of tube atoms, and the factor 2 in front of the sum is introduced to account for the number of interacting surface atoms. The latter is approximated to be identical to the number of tube interacting atoms due to the similar surface atom densities that they exhibit. For comparison purpose, we present the LJ energy variations of Fig. 2b normalized using the same procedure. The main differences observed when switching to the KC force field are: (i) increased sliding energy corrugation[26] and (ii) decreased weight of the smaller peak appearing at a sliding distance of 5.7 Å. The former can be accounted for by replacing the RI scaling factor of 1.97 meV/atom appropriate for the LJ sliding corrugation by 6.12 meV/atom for the KC case. The latter results from the anisotropic contribution introduced in the KC potential and requires a slight modification of the RI definition in order to be captured. To this end, the orientation of the radial $p$ orbital associated with each tube or surface atom is defined as the normal to the surface formed by its nearest-neighboring sites. For each pair of atoms, one residing on the tube and another on the surface, two lateral distances are calculated between the tube atom and the normal to the surface atom and between the surface atom and the normal to the tube atom. The corresponding pair Gaussian overlap is then defined as the average of Gaussian overlaps calculated using the two lateral distances (for more details see the supplementary material). As can be seen, the RI is able to reproduce both the LJ and KC traces thus demonstrating its robustness.

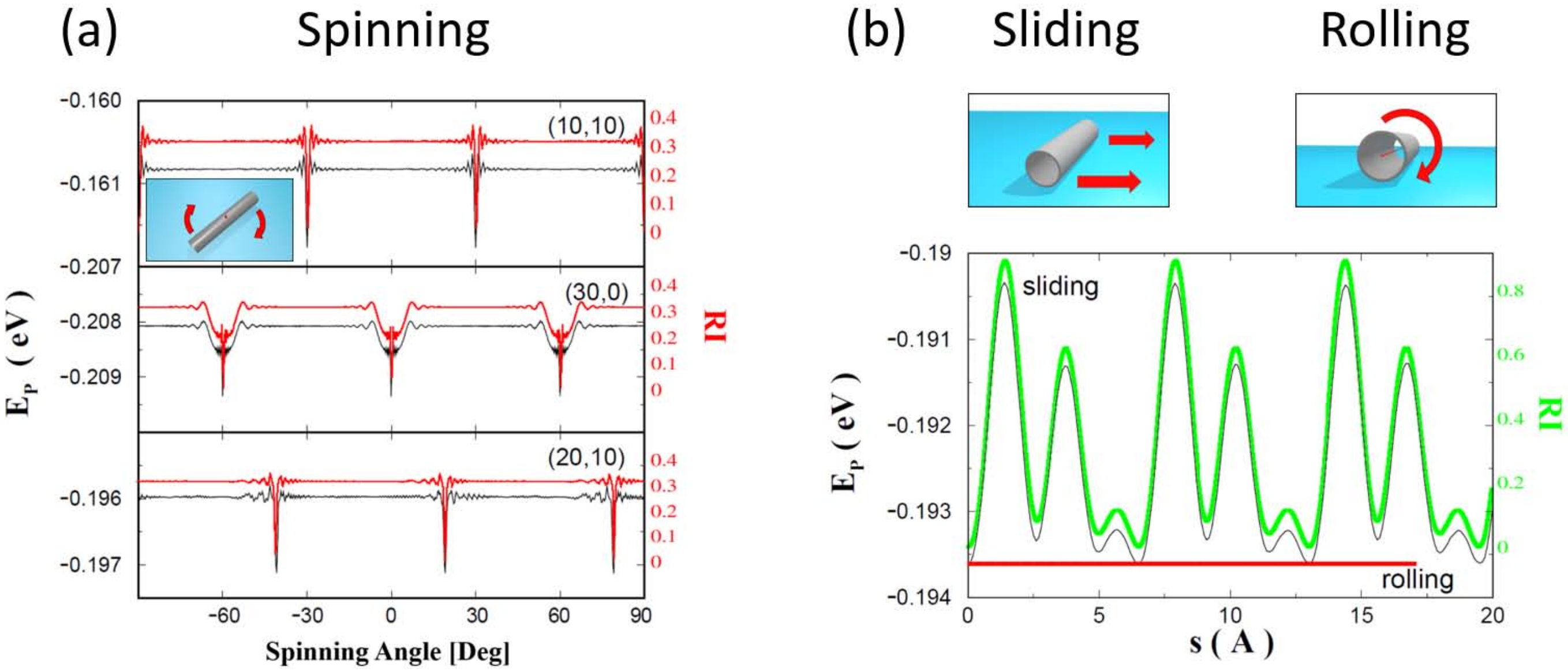

Figure 2: (a) Lennard-Jones energy (black) and RI (red) variations as a function of the spinning angle for $(10, 10)$, $(30, 0)$, and $(20, 10)$ CNTs on graphene (upper, middle, and lower panels, respectively). The RI calculations, performed with NT lengths of 10 nm, 30 nm, and 34 nm for the $(10, 10)$, $(30, 0)$, and $(20, 10)$, respectively, are slightly shifted upwards for clarity. Illustration of the spinning motion around the perpendicular axis is presented in the inset of the upper panel. (b) Lennard-Jones energy (black) and RI variations as a function of the sliding (green) and rolling (red) distance for a $(20, 10)$ CNT on graphene. For clarity, the RI sliding curve is slightly shifted upward. Illustration of the sliding and rolling motions are provided in the upper left and right insets, respectively. The 2D Gaussian standard deviation for the carbon atoms was taken as $\sigma_C = 0.375L_{CC}$. Sample coordinates for each system are provided in the supplementary material. The Lennard-Jones reference results were adopted with permission from A. Buldum and Jian Ping Lu., *Phys. Rev. Lett.* **83,** 5050 (1999).[13] Copyright (1999) by the American Physical Society.

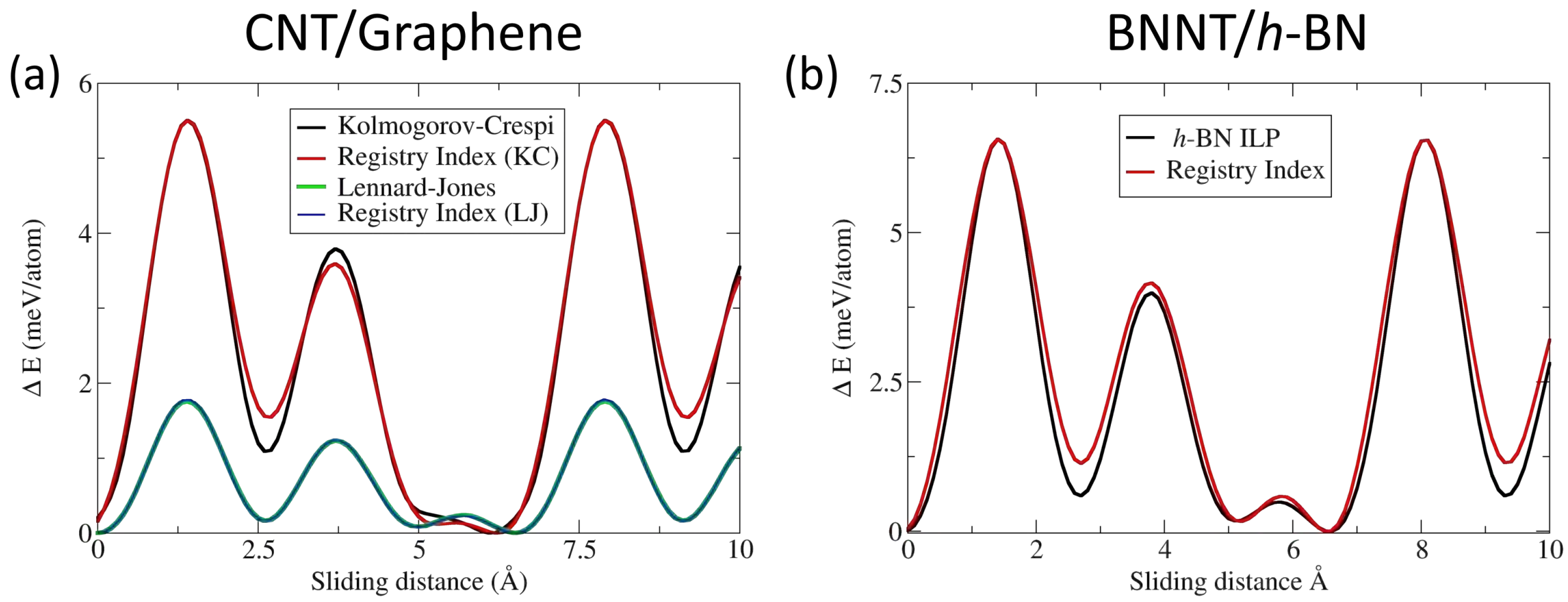

Figure 3: Nanotube sliding energy landscapes calculated using anisotropic interlayer potentials. (a) KC sliding energy (black) and RI (red) variations of a $(20, 10)$ CNT of length 57 nm as a function of the sliding distance. The LJ landscape (green) and the corresponding RI trace (blue) are presented for comparison purposes. (b) $h$-BN ILP sliding energy (black) and RI (red) landscapes of a $(20, 10)$ BNNT of length 57 nm as a function of the sliding distances. In the initial configuration the CNT (BNNT) atoms closest to the surface are positioned at the AB ($AB_1$) stacking mode of graphene ($h$-BN) see Fig. 1a. The 2D Gaussian standard deviations used in these calculations are $\sigma_C = 0.375L_{CC}$, $\sigma_B = 0.1125L_{BN}$ and $\sigma_N = 0.375L_{BN}$ for the carbon, boron, and nitrogen atoms, respectively, where $L_{BN} = 1.45$ Å is the covalent boron-nitrogen bond length in $h$-BN. The RI results are multiplied by the appropriate scaling factors and vertically shifted to match the force-field diagrams.

Next, we turn to study the motion of a BNNT on a flat $h$-BN surface. To this end, we utilize our recently developed $h$-BN interlayer potential[16] as a benchmark for the RI calculations. Similar to the graphitic case studied above, we consider the sliding motion of a $(20, 10)$ BNNT on $h$-BN. In Fig. 3b we plot the force-field (black) and RI (red) variations during the sliding motion. The BN system shows general features that resemble those of the graphitic junction with a three-peaks periodic structure. We note that the position of the peaks is slightly shifted due to the longer BN covalent bond leading to a larger tube diameter. Here, as well, the RI successfully reproduces the force field results, with a scaling factor of 8.67 meV/atom. Notably, this good agreement is obtained with an overlap downscale rate identical to the one used for modeling the graphitic systems ($\alpha_{h-\mathrm{BN}} = \alpha_g$). This indicates that the dominating interactions are associated with the surface-facing tube atoms resulting in weak dependence of the results on the choice of overlap downscale rate as discussed above.

Having validated the ability of the RI to describe the energy landscape of nanotube motion on flat hexagonal surfaces of homogeneous systems we may now use it to predict the behavior of the heterogeneous junctions of CNT/$h$-BN and BNNT/graphene. As before, we consider a 57 nm long $(20, 10)$ nanotube initially positioned as depicted in Fig. 1a. The corresponding RI variations are presented by the green and blue curves in fig. 4 for the CNT/$h$-BN and BNNT/graphene junctions, respectively. For comparison purpose we also present the corresponding results for the homogeneous CNT/graphene (black) and BNNT/$h$-BN (red) systems. Due to the in-plane lattice vectors mismatch between graphene and $h$-BN the hetero-junctions exhibit significantly reduced RI variations. We note that care should be taken when comparing the RI landscapes of the various junctions as they require different scaling factors to capture the calculated sliding energy variations. Nevertheless, since the scaling factors of the homogeneous systems differ by less than 30% and since the corresponding scaling factors of the heterogeneous junctions are expected to be similar or lower (due to the intrinsic lattice vectors mismatch) we can deduce that the latter will present a much less corrugated sliding energy landscape. Therefore, assuming that the interface is clean and that the motion is wearless one may expect lower friction or even superlubric behavior of the rigid heterogeneous junctions.

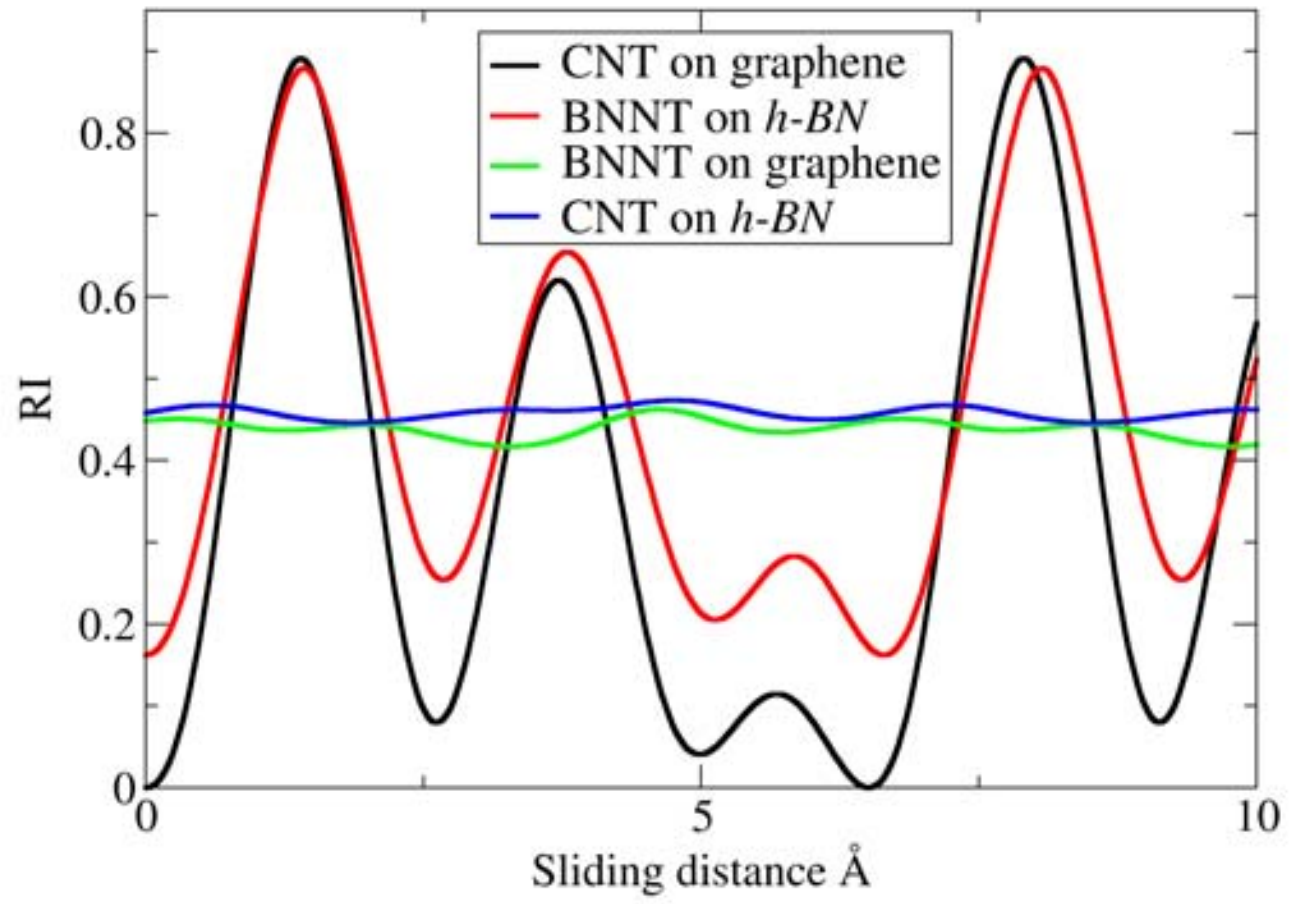

Figure 4: Comparison of sliding energy landscapes for homogeneous and heterogeneous interfaces of 57 nm long $(20, 10)$ nanotubes and hexagonal surfaces. The tubes are initially positioned at the AB stacking mode with the main axes aligned at an angle of $30° - \theta$ with the zigzag direction (see supplementary material for sample coordinates). The 2D Gaussian standard deviations of the homogeneous interfaces was taken as $\sigma_C = 0.375L_{CC}$, $\sigma_B = 0.1125L_{BN}$, and $\sigma_N = 0.375L_{BN}$, for the carbon, boron, and nitrogen atoms, respectively. For the heterogeneous junctions the corresponding standard deviations used are $\sigma_C = 0.375L_{CC}$, $\sigma_B = 0.15L_{BN}$, and $\sigma_N = 0.3L_{BN}$.

# Summary and Outlook

In this study, the applicability of the Registry Index method has been extended to the realm of non-parallel rigid interfaces. By rescaling the overlap terms according to the corresponding inter-atomic distance we were able to capture the energy variations during the sliding, rolling, and spinning motions of CNTs on graphene calculated using both the Lennard-Jones and Kolmogorov-Crespi classical force-fields. Furthermore, the generalized RI was able to reproduce the sliding energy landscape of a BNNT on $h$-BN calculated using the $h$-BN ILP. In order to model anisotropic interactions introduced by the KC and $h$-BN ILP potentials the calculations of the inter–atomic distance entering the RI overlaps explicitly involve the local surface normal vectors. Furthermore, the circle associated with each atom in the original RI implementation were replaced by Gaussians to allow for smoother and more physical RI landscapes at no extra computational cost. Finally, the sliding behavior of a CNT on $h$-BN and a BNNT on graphene were studied using the developed approach indicating that superlubric behavior may be expected to occur at such heterogeneous interfaces.

## Acknowledgments

We would like to thank Dr. Yael Roichman for her help with some of the graphics in this paper. This work was supported by the Lise-Meitner Minerva Center for Computational Quantum Chemistry and the Center for Nanoscience and Nanotechnology at Tel-Aviv University. I.O. would like to thank Ofer Horowitz for helpful discussions.

# Nanotubes Motion on Layered Materials: A Registry Perspective

## Supplementary Information

Inbal Oz,[1,2] Itai Leven,[1,2] Yaron Itkin,[1] Asaf Buchwalter,[1] Katherine Akulov,[1] Oded Hod[1,2]

[1]Department of Physical Chemistry, School of Chemistry, The Raymond and Beverly Sackler Faculty of Exact Sciences, Tel Aviv University, Tel Aviv, IL 6997801

[2]The Sackler Center for Computational Molecular and Materials Science, Tel Aviv University, Tel Aviv, IL 6997801

**Replacing atomic centered circle overlaps by projected Gaussian overlaps**

In the original implementation of the registry index method atomic centered circles were used to evaluate the degree of inter-lattice commensurability at rigid interfaces. This choice allowed the efficient evaluation of projected circle overlaps via a simple analytic expression. Notably, this simplistic picture was able to rationalize the measured frictional behavior of a bilayer graphene junction.[1,2] Nevertheless, the obtained RI surfaces were characterized by somewhat sharp features when compared to the corresponding sliding energy landscapes calculated using advanced first-principles method.[3] In order to obtain smoother and more physical RI surfaces that better match the reference calculations we have replaced the atomic centered circles by two-dimensional Gaussians. The latter provide a softer reduction of the projected pair overlaps as a function of lateral inter-atomic distance (see Fig. S1) while maintaining the analytic nature of the overlap area evaluation.

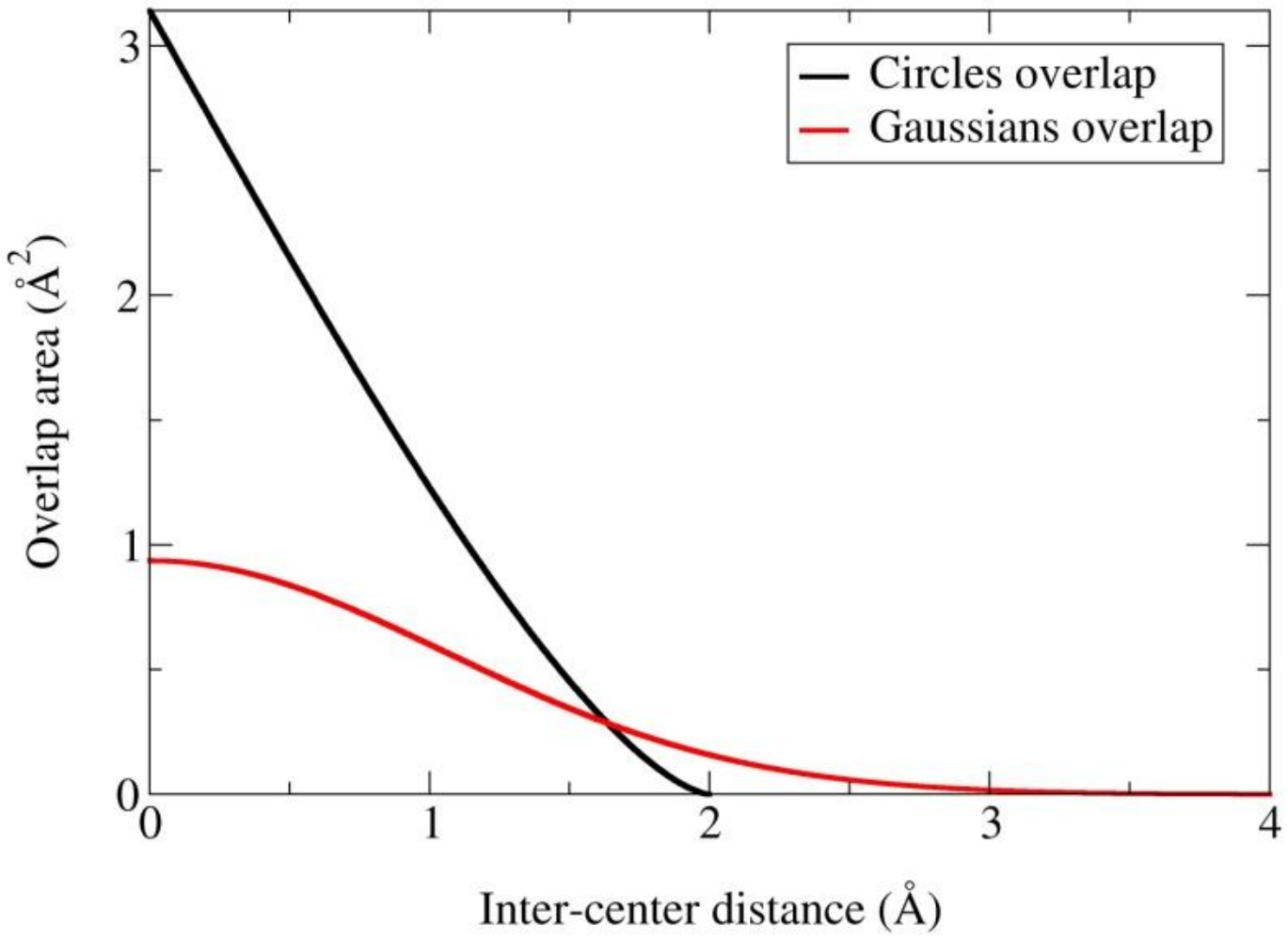

**Figure S1: Circle vs. Gaussian pair overlap.** Comparison of the pair overlap of two circles (black) and two Gaussians (red) as a function of their inter-center distance, d. The pair circle overlaps are calculated using $S_{ij}^{circle} = \mathrm{r_i^2} \cos^{-1}\left[\left(\mathrm{d}^2 + \mathrm{r_i^2} - \mathrm{r_j^2}\right)/\left(2\mathrm{dr_i}\right)\right]$ $+\mathrm{r_j^2}\cos^{-1}\left[\left(\mathrm{d}^2 + \mathrm{r_j^2} - \mathrm{r_i^2}\right)/\left(2\mathrm{dr_j}\right)\right] - 0.5\sqrt{\left(\mathrm{r_i} + \mathrm{r_j} - d\right)\left(\mathrm{r_i} - \mathrm{r_j} + d\right)\left(\mathrm{r_j} - \mathrm{r_i} + d\right)\left(\mathrm{r_i} + \mathrm{r_j} + d\right)}$, where $\mathrm{r_i} = \mathrm{r_j} = 1$ Å are the two circle radii. The pair Gaussian overlaps are calculated using $S_{ij}^{Gaussian} = \left(\pi\sigma_i^2\sigma_j^2 / \sqrt{\sigma_i^2 + \sigma_j^2}\right) e^{-0.5d^2/\left(\sigma_i^2+\sigma_j^2\right)}$ with $\sigma_i = \sigma_j = 0.75$ Å being the Gaussian standard deviations.

To demonstrate the performance of the new implementation we compare, in Fig. S2, the sliding RI surface of periodic bilayer *h*-BN obtained using circle (left panel) and Gaussian (middle panel) overlaps with the corresponding sliding energy landscape (right panel) obtained using the TS-vdW dispersion corrected PBE exchange-correlation density functional approximation with the tier-2 basis set of the FHI-AIMS code given in Ref. [3]. As can be seen, the RI landscape obtained using Gaussian overlaps is in good agreement with the reference DFT results providing a better physical picture than the corresponding circle overlap based calculation without any increase in computational cost or complexity.

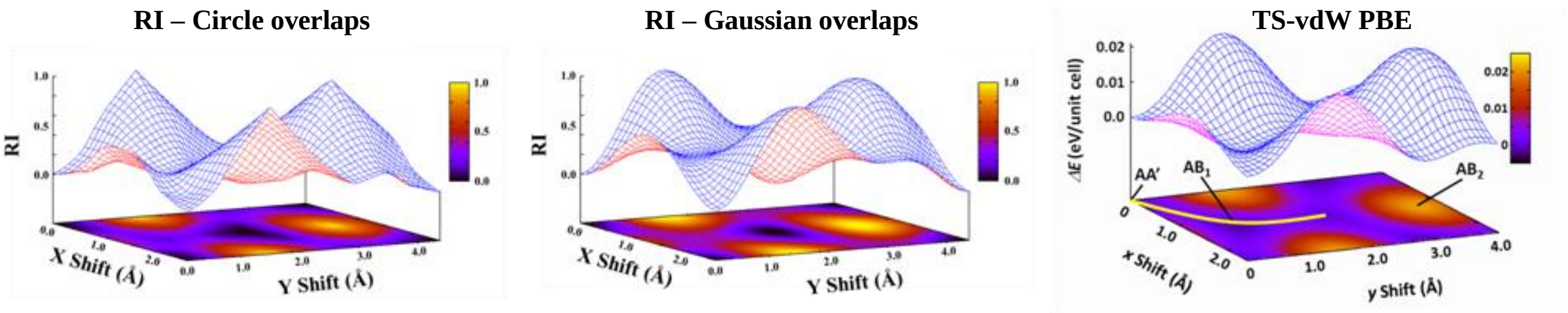

**Figure S2: Circle vs. Gaussian overlap sliding RI landscapes.** Comparison of sliding RI landscapes obtained using circle overlaps (left panel) and Gaussian overlaps (middle panel) with the reference sliding energy calculations of bilayer $h$-BN (right panel) obtained using the TS-vdW dispersion corrected PBE exchange-correlation density functional approximation and the tier-2 basis set as implemented in the FHI-AIMS code.[3] The circle radii used are $\mathrm{r_N} = 0.5\mathrm{a_{BN}}$ and $\mathrm{r_B} = 0.15\mathrm{a_{BN}}$ with $\mathrm{a_{BN}} = 1.446$ Å being the equilibrium BN bond length in $h$-BN. Gaussian standard deviations are given by $\sigma_N = 0.75\mathrm{r_N}$ and $\sigma_B = 0.75\mathrm{r_B}$.

## Anisotropic pair-overlap registry index calculation

When calculating the nanotube/substrate atomic centered Gaussian overlaps within the registry index method curvature effects should be taken into account. To do so we follow the Kolmogorov-Crespi[4] and $h$-BN ILP[5] approach. In these methods the inter-layer interactions depend on the lateral distance $\rho_{in}$ (see Fig. S3) between atoms $i$ and $n$ on adjacent layers. $\rho_{in}$ is defined as the distance between atom $n$ of one layer and the surface normal at the position of atom $i$ of the other layer. The latter is taken as the normal to the surface defined by the three nearest-neighbors of atom $i$ within the hexagonal lattice.

In the original implementation of the registry index method for flat parallel surfaces, the circle overlaps were calculated as a function of the lateral distance $\rho_{in} = \rho_{ni}$ between each pair of atoms ($i$ and $n$) in adjacent layers. When considering non-parallel surfaces the lateral distance is no longer symmetric such that $\rho_{in} \neq \rho_{ni}$. To account for this, the Gaussian pair overlap is taken as the average of the overlap calculated using $\rho_{in}$ and $\rho_{ni}$ in the following manner:

$$S_{in} = 0.5 \cdot [S(\rho_{in}) + S(\rho_{ni})], \tag{1}$$

where $S(\rho_{in})$ and $S(\rho_{ni})$ are the Gaussian overlaps of atoms $i$ and $n$ calculated using Eq. 1 of the main text at lateral distances $\rho_{in}$ and $\rho_{ni}$, respectively, and $S_{in}$ is the average overlap area of the atomic pair. With this definition the overlap becomes dependent on the relative spatial orientation of the two 2D-Gaussians and reduces with increasing inter-Gaussian tilt angle.

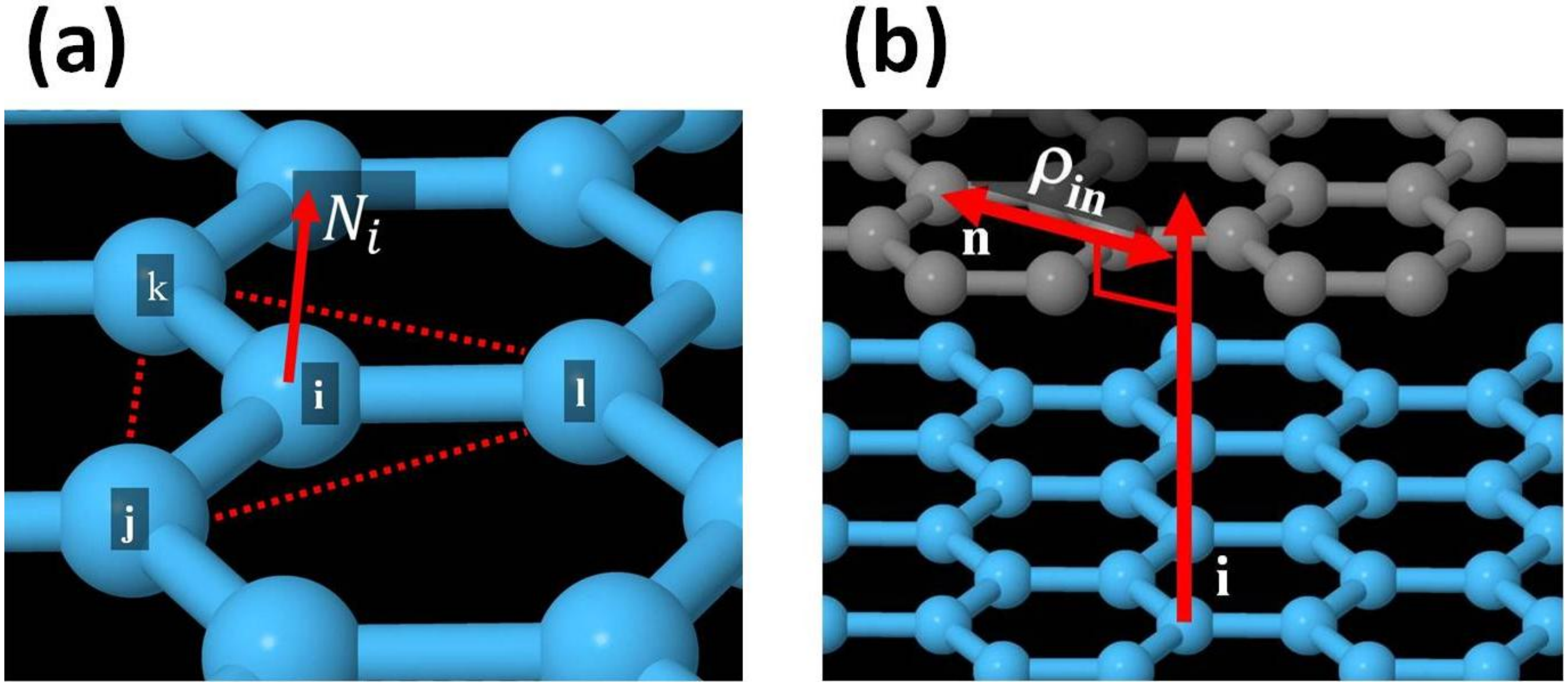

**Figure S3: Normal and lateral inter-atomic distance definitions.** (a) The surface normal at the position of atom $i$, $N_i$, is defined as the normal to the plane defined by its three nearest-neighbors $j$, $k$ and $l$. (b) The lateral distance $\rho_{in}$ is the shortest distance between atom $n$ of one layer and the normal of atom $i$. For clarity we show here a bilayer graphene system and color the lower layer atoms in cyan.